\documentstyle[11pt,paspconf,epsf]{article}

\begin{document}

\title{The Nature of the Compact/Symmetric Near-IR Continuum Source in
4C 40.36} 
\author{E. Egami\altaffilmark{1},
L. Armus\altaffilmark{2}, G. Neugebauer\altaffilmark{1},
B. T. Soifer\altaffilmark{1}, A. S. Evans\altaffilmark{3}, and
T. W. Murphy Jr.\altaffilmark{1}} \affil{1. 320-47 Downs Laboratory of
Physics, California Institute of Technology, Pasadena, CA 91125}
\affil{2. SIRTF Science Center, IPAC 100-22, California Institute of
Technology, Pasadena, CA 91125} \affil{3. Department of Physics and
Astronomy, State University of New York at Stony Brook, Stony Brook,
NY 11794-3800}

\begin{abstract}
Using NICMOS on HST, we have imaged the emission-line nebulae and the
line-free continuum in 4C 40.36, a ultra-steep spectrum FR II radio
galaxy at $z=2.269$.  The line-free continuum was found to be
extremely compact and symmetric while the emission-line nebulae seen
in H$\alpha$+[N II] show very clumpy structures spreading almost
linearly over 16 kpc.  However, this linear structure is clearly
misaligned from the radio axis.  The SED of the line-free continuum is
very flat ($f_{\nu} \propto \nu^{-0.5}$), suggesting that if the
continuum emission is produced by a single source, it is likely to be
a young bursting stellar population or scattered AGN light.  However,
because of the lack of a line-free optical image with a comparable
spatial resolution, we cannot exclude the possibility that the
observed SED is a composite of a young blue population and an old red
population.
\end{abstract}


\section{Introduction}
The compact/symmetric morphology of high-$z$ radio galaxies seen in
the near-IR is often taken as the evidence that radio galaxies contain
an old, dynamically relaxed stellar population, whose slow passive
evolution is supposed to maintain the tight $K-z$ relation over a wide
redshift range (Lilly \& Longair 1984; Lilly 1989; Rigler et
al. 1992).  A consequence of this picture is that radio galaxies are
old, and that the bulk of their mass must have been assembled at high
redshifts.  

On the other hand, near-IR spectroscopic studies of high-$z$ radio
galaxies have shown that when the near-IR broad-band magnitudes are
corrected for strong line emissions, the continuum color of some
high-$z$ radio galaxies becomes extremely blue, which prompted an
interpretation that some of these galaxies might be extremely young
protogalaxies (Eales \& Rawlings 1993).

Here, we present high-resolution HST/NICMOS images of 4C 40.36, an
ultra-steep spectrum FR II radio galaxy at $z=2.269$ (Chambers, Miley,
\& van Breugel 1988).  In addition, we also present a low-resolution
($R \sim 80$) $H$+$K$ band spectrum of 4C 40.36 taken with Keck/NIRC.
By combining the imaging data and spectrum, we are able to estimate
the overall line contribution to the $H$- and $K$-band fluxes, and
obtain the morphology of the emission-line gas and the continuum in 4C
40.36.

\begin{figure}[t]
  \vspace*{-1.cm}

  \plotone{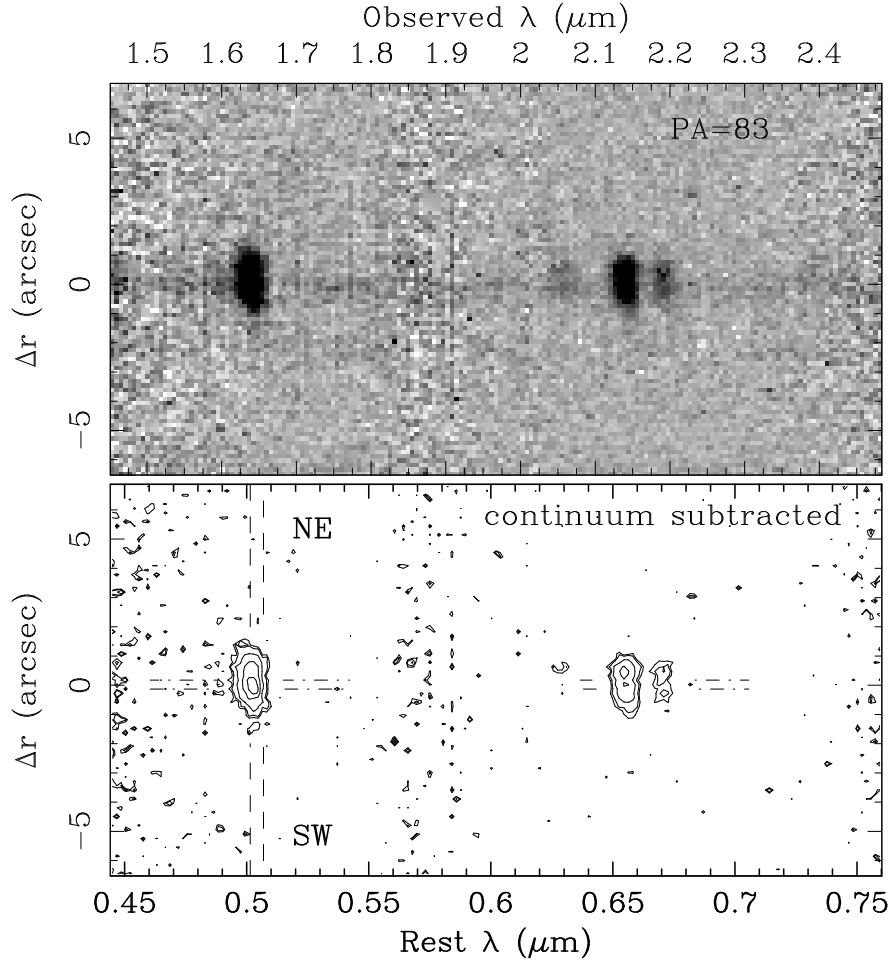} 
  \vspace*{-5.3cm}
  \caption{A near-IR NIRC low-resolution ($R \sim 80$) grism spectrum
  of 4C 40.36.  The PA of the slit corresponds to that of the radio
  axis (83$^{\circ}$ E of N, shown in Figure~2).  The slit width is
  0\farcs6.  The emission lines visible in the spectrum are [O~III],
  [O~I], H$\alpha+$[N~II], and [S~II] from left to right.}

\end{figure}

\section{The Strong Emission Lines and Continuum Color of 4C 40.36}

\begin{figure}
\plotone{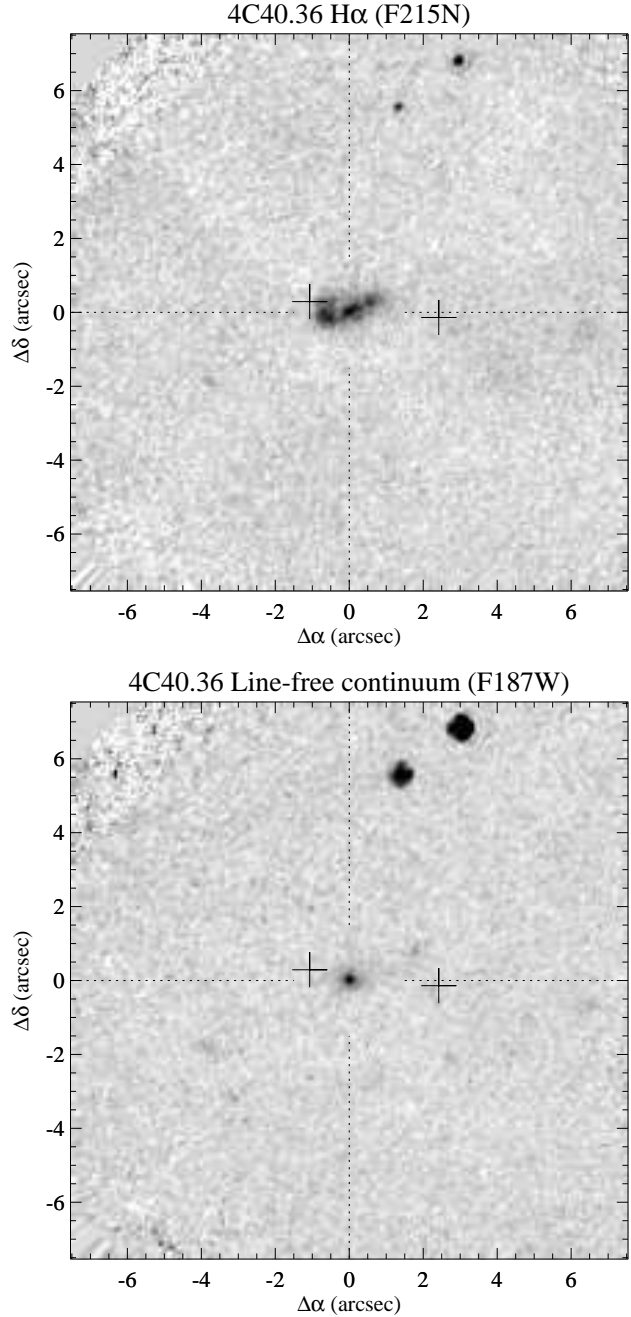}

\vspace{-0.5cm}

\caption{Upper: A narrow-band H$\alpha$ image of 4C 40.36 taken with
the NIC 2 camera and the F215N narrow-band filter.  The pixel scale is
0.075\arcsec/pixel.  The crosses indicate the positions of the radio
lobes; Lower: A line-free continuum image of 4C 40.36 taken with the
NIC 2 camera and the F187W wide-band filter.  The core is unresolved
(PSF FWHM $=$ 2.6 pix (0.2\arcsec) $=$ 1.6 kpc measured for the two
stars near the top), but there is a faint circularly symmetric halo
around it.}
\end{figure}

Since a number of strong restframe-optical emission lines are
redshifted into the near-IR bands at this redshift, the flux
contribution from these lines must be first evaluated.  The Keck/NIRC
spectrum clearly shows the strong [O III], H$\alpha$+[N II], and [S
II] lines (Figure~1).  From this spectrum, we estimate that the line
contribution is 60 \% in the $H$ band and 70 \% in the $K$ band,
respectively.

The dominant emission line in the $K$ band is H$\alpha$+[N II], and
the NICMOS narrow-band image of the H$\alpha$+[N II]-emitting nebulae
is shown in Figure~2 (upper panel).  It is extended almost linearly
over 2\arcsec ($\sim$ 16 kpc; $q_{0}=0.5$, H$_{0}=50$ km s$^{-1}$
Mpc$^{-1}$) with three bright knots.  The eastern region shows a
significant extension going up to the north.  We also note that this
H$\alpha$+[N II] linear structure is misaligned from the radio axis,
which indicates that it is not the expanding jet that is ionizing the
gas.

The line-free continuum source was found to be extremely compact and
symmetric (Figure~2, lower panel).  It has an unresolved core (PSF FWHM
$=$ 0\farcs2 $=$ 1.6 kpc), and an almost circularly symmetric halo
extending out to at least a radius of $\sim$ 4 kpc.

Based on the line-free continuum images at 1.45 $\mu$m (not shown
here) and 1.87 $\mu$m, we conclude that the continuum SED is flat
($f_{\nu} \sim$ const.) at $\sim$ 10 $\mu$Jy within this range.  Even
after the line-free near-IR magnitudes are combined with the line-free
$V$ and $R$ magnitudes derived by Chambers, Miley, \& van Breugel
(1988), the overall SED remains very flat ($f_{\nu} \propto
\nu^{-0.5}$).

\section{Interpretation}
In the case of 4C 40.36, the extreme blue color of the continuum seems
to suggest that the source is a young bursting stellar population
possibly forming a spheroid or scattered AGN light (e.g., Eales \&
Rawlings 1993; Iwamuro et al. 1996).  On the other hand, we cannot yet
exclude the possibility that the optical continuum source and the
near-IR continuum source are two completely separate components.  In
fact, a morphological study of $z\sim1$ radio galaxies suggests that
radio galaxies seem to have both a red old stellar component, which is
compact, symmetric, and provides most of the mass, and a blue young
stellar component, which is luminous and often aligned with the radio
axis, but involves only a small amount of mass (Rigler 1992).  If this
is the case, the blue color of the integrated total flux does not
necessarily mean that 4C 40.36 is a young galaxy.  To remove this
ambiguity, it is necessary to obtain a line-free optical (i.e.,
restframe UV) image with a spatial resolution comparable to that of
HST/NICMOS.


\begin{references}
\reference Chambers, K. C., Miley, G. K., \& van Breugel,
W. J. M. 1988, \apj, 327, L47
\reference Eales, S. A., \& Rawlings, S. 1993, \apj, 411, 67
\reference Iwamura, F., Oya, S., Tsukamoto, H., \& Maihara, T. 1996,
\apjl, 466, L67
\reference Lilly, S. J. 1989, \apj, 340, 77
\reference Lilly, S. J., \& Longair, M. S. 1984, \mnras, 211, 833
\reference Rigler, M. A., Lilly, S. J., Stockton, A., Hammer, F., \& Le
Fevre, O. 1992, \apj, 385, 61
\end{references}
\end{document}